\documentclass[angeod,manuscript]{copernicus}  

\nolinenumbers
\usepackage{amsmath}
\usepackage{color}

\usepackage{natbib}

\begin{document}

\title{{Auroral Kilometric Radiation -- the Electron Cyclotron Maser Paradigm
}}

\author[1]{W. Baumjohann}
\author[3]{R. A. Treumann}
\affil[1]{Space Research Institute, Austrian Academy of Sciences, Graz, Austria\\
\affil[2]{International space Science institute, Bern, Switzerland}
\emph{Correspondence to}: Wolfgang.Baumjohann@oeaw.ac.at
}

\runningtitle{AKR and ECMI}

\runningauthor{Baumjohann and Treumann}

\received{ }
\pubdiscuss{ } 
\revised{ }
\accepted{ }
\published{ }


\firstpage{1}

\maketitle

\begin{abstract}
Auroral kilometric radiation (AKR) is the paradigm of intense radio emission from planetary magnetospheres. Being close to the electron gyro frequency and/or its lower harmonics, its observation indicates the non-thermal state of the source plasma. Emission is produced when the plasma enters a state of energetic excitation which results in deformation of the electron distribution function. Under certain conditions this leads to "quasi-coherent" emission. It is believed that the weakly-relativistic electron-cyclotron-maser instability is responsible for this kind of radiation. Since energetically radio radiation normally is  not of {primary} importance in the large-scale magnetospheric phenomena, AKR as such has, for the purposes of large-scale magnetospheric physics, become considered a marginal problem. Here this notion is questioned. AKR while applying to the auroral region mainly during magnetospherically disturbed times {carries just a fraction of the total substorm energy. It is, however, of diagnostic power in the physics of the upper auroral ionosphere and Space Weather research}. As a fundamental physical problem of generation of radiation in non-thermal plasmas {it remains not resolved yet. Many questions have been left open even when dealing only with the electron-cyclotron-maser. These can advantageously be studied in the magnetosphere proper both by observation and theory, the only continuously accessible place in space. The most important are listet here with hint on how they should be attacked. Its value is to be sought in the role it should play in application to the other magnetized planets, extra-solar planets, and to  strongly magnetized astronomical objects} as an important {tool to diagnose the matter state responsible for radiation in the radio frequency range beyond thermal, shock or synchrotron radiation}.
    

\end{abstract}
\section{Introduction}
Attaching the label \emph{solved} to any field in science is, to some extent, dangerous;  the least to say, it is  problematic. It may readily discourage any researcher to further pursue this field.
Contemplating about which {particular} problems in magnetospheric physics might have been solved in the past, it is hard to escape the impression that, in reality, {there might be very few only or even none at all}. The {obvious} reason is that any problem that has been brought to some preliminary solution, always generates an entirely new set of related problems which emerge from the solution found. Worse still, if a problem would really be solved ultimately, then the whole field has come to an end. It has become dead and disappears from science as a problem of no scientific interest anymore, with the exception of history or/and its technical application. But this happens very rarely only.

As for a well known example, Newton's law of gravitation was for about two centuries believed to be ultimate wisdom which nobody was considering of any interest except in (very successful) application to planetary, astronomical, and engineering problems, even though Newton himself was worried about its origin. Science took it for granted and had put it on the shelf of solved problems. This held until a few people sharing Newton's doubts  started digging into its relevance, notably Ernst Mach who contemplated about its origin grounded in the (undeniable) existence of the entire universe, and Roland E\"otv\"os who performed his famous experiment on the mass equivalence that led to Einstein's equivalence principle (which the MICROSCOPE satellite has just confirmed again with unprecedented precision \citep{touboul2022}) and the emergence of General Relativity, still without solving the mystery of the existence of mass and the origin of gravitation other than in the curvature of space, which was a circular argument and thus remained unresolved. 

Another example is Maxwell's ingenious formulation of the laws of Electrodynamics which after the experimental proof of electromagnetic waves by Heinrich Hertz and his insight that light is but electromagnetic waves was believed to be complete and ready for endless application, a notion that had readily become vindicated by Einstein's Special Relativity and his insight into the particle nature of light which was understood though not explained only after the formulation of Quantum Electrodynamics. This list can be extended. Nevertheless, considering just special problems within one bigger scientific theory, one may speak of \emph{partially solved problems}, those that hold within some approximations made, the boundaries within which one must not anymore worry about its applicational limitations.

{Ultimately solving any problem, seems to be inhibited in Nature; if it really happens then, with very high probability, the solution was either wrong (like Aristoteles' world view, Ptolemy's universe,  Phlogiston theory, violation of energy conservation, existence of any ghostly non-physical life-force, creation theory, and other more modern candidates we cautiously do not mention here \footnote{Including any philosophical world view. History demonstrated that, without any exception, they have all been wrong and, when practiced, ending up in ideology and causing severe damage to human civilisation.}) or incomplete (like black hole theory remained until the discovery of black hole entropy, temperature, and radiation).} 

Magnetospheric or, to a larger extent, Space Plasma Physics is kind of such a limited field in the realm of the more general Space and Astrophysics. {Its gross physics on the planetary scale (magnetoshere as obstacle in the solar wind, shape, global structure, etc.) is to some extent known and could be qualified as partially solved. It forms the macroscopic basis of Space Weather (a better name would be Space Meteorology). Already for long it is fighting with its perfection which, similar to ordinary Meteorology, will with very high probability never ever be achieved for the same reason: the complexity of processes it is dealing with.}

There are more subtle magnetospheric effects which would qualify to be called \emph{partially} solved, like the limitation of high energy particle fluxes in the radiation belts which have been treated in the midst of the past century in a landmark paper \citep{kennel1966} applying the then in plasma physics new quasilinear theory in relation to atmospheric nuclear tests,  demonstrating the relevance of wave-particle interactions in causing observable macroscopic effects. Several problems concerning the formation and structure of the bow shock have also been partially solved: the almost trivial subdivision of the shock into quasi-perpendicular and quasi-parallel is, within its large bounds, a quasi-rigid solution, as rigid as the existence of the shock itself, including its effect on structuring the shock, its surrounding particle and wave distributions, and the existence of the downstream quasi-thermalized though turbulent transition region, the magnetosheath. In fact the physics of shock and magnetosheath are under scrutiny and in fast evolution; the physics of the magnetopause struggles with the  observationally and theoretically still completely unsolved problem of reconnection. In hindsight they all generated many surprises, troubles, and an extraordinarily large number of publications, becoming sources of further research in unpredictable directions.

Let us stop here. Of the excessive number of remaining problems we here pick one: 
the generation of auroral kilometric radiation (AKR). {We focus  on} the theoretical understanding of its generation mechanism, {providing a rather cursory account only of some of the observational facts. Today AKR is believed to be generated by the electron cyclotron maser instability (ECMI) which, in our view justifiably, has become canonical. We nevertheless identify a number of remaining open questions which to us seem important and in the future should be addressed in four ways: by developing sufficiently high resolution  instrumentation to fly on spacecraft; observations/measurements  inside the AKR source region (which have been disregarded during the few last decades), and also on the ground, in order to infer its propagation, amplification/absorption/damping, transformation, and transport; by statistical methods and, which interests us here most, theory. We will also highlight some of the general importance AKR might/should have in astrophysics.}

\section{Briefing on observations of AKR}

\begin{figure*}[t!]
\centerline{\includegraphics[width=0.5\textwidth,clip=]{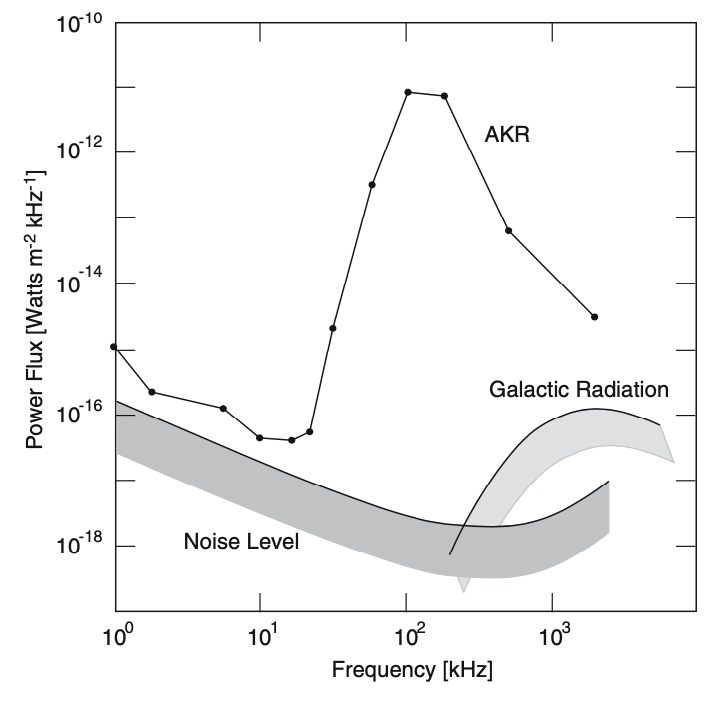}}
\caption{The spectral intensity of Gurnett's  \citep{gurnett1974} first observation and identification of auroral kilometric radiation (adapted in this form from \citep{treumann2006}).} \label{fig1}
\end{figure*}

That the inhabited earth radiates radio signals into space is a trivial fact. This radiation is the result of the many artificial transmitters and transmitter stations on the earth. It is narrow band and sporadic, most of it is directed, and all of it decays rapidly with radial distance from earth \citep{labelle1989}, causing pollution of near-earth space by anthropogenic radio noise.  ``Extraterrestrials'' (Setis, who do not exist near earth) could use it as unambiguous  indication of intelligent life on earth \citep{sagan1993}. Hence it was no little surprise when in the mid-sixties the Russians with Elektron-2, one of their satellites \citep{benediktov1965} observed a narrow-band natural radio emission near earth which differed from any artificial radiation. Its origin and distribution could not be identified at that time. One decade later this observation was confirmed {with the Imp 6 and 8 spacecraft} and made precise in another landmark paper \citep{gurnett1974} which determined the approximate radiation center frequency near 200 to 300 kHz ({with spectral power density shown in Fig. \ref{fig1}}), bandwidth to few 100 kHz, approximate source location, {determined from one year of continuous observations}, in the nightside upper auroral magnetosphere above several 1000 km altitude \citep{kurth1975}, presumably during auroral substorms \citep{benson1984} (the major macroscopic magnetospheric disturbance), {though not continuously present, directed outward and highly beamed. Its approximate integrated intensity was estimated from Imp-6 at geocentric distance $r\approx 30$ R (earth radii) and solid beaming angle $\approx6.5$ ster to $\sim 10^9$ W (restricting to the most intense events) which was later rectified, based on better statistics and including much weaker events, to the interval $10^7-10^9$ W} \citep{green1977,green1979,green1985,omidi1984,kurth1998}, depending on bandwidth and duration. The moderate and intense emissions  (now dubbed auroral kilometric \citep{kurth1975}) occur during  magnetospheric substorms above the disturbed auroral ionosphere. { Comparing the numbers to the nominal average substorm power (determined from the strength of the auroral currents, assuming a $\sim 30$ min canonical substorm duration) of $\sim 10^{11}$ W  \citep{kamide1993,ieda1998} this range amounts to $(0.01-1)$\% of substorm energy -- an unexpectedly large fraction for any radiation in the radio band!}  {The outward beaming of the radiation clearly points to its origin near earth in the upper auroral magnetosphere. Being highly perpendicular to the magnetic field thus propagating mostly in the X mode polarization as has been convincingly proved by ray path calculations and observation \citep{omidi1984}. The original claim that its source region coincides just with the ring current in the late evening magnetosphere is in general vindicated. Its sources are more localized than originally believed. In particular weaker events are frequent and distributed over the entire nightside auroral zone from dawn to dusk. These regions glow  in weak AKR even under only moderately disturbed times \citep{morioka2005}.They are even found on the dayside under the cusp. It remains to be unclear whether AKR is excited in the upward or downward current region, or both. }

Mainly for its high power, in the years following \citep{gurnett1974} interest in AKR became substantial. {Almost every spacecraft that carried a wave instrument in the appropriate spectral range and flying on non-equatorial orbit attempted to contribute to observation at varying distance from earth and magnetic inclination. The complete list of those spacecraft would be long, including even those which were devoted to observations in the more distant tail, solar wind, other planetary environments, and the heliosphere; it also extends into the present. We find it appropriate to provide a \emph{selected} list of the spacecraft which provided the observational base in Table \ref{list}, ordered into two groups, those with low and those with high inclination, the latter orbiting on approximately polar orbits and low apogee. (An about complete list is given in \citep{fogg2022,waters2022} who very recently used Wind spacecraft observations of AKR to nicely infer about solar wind-magnetosphere coupling in view of space weather applications.) Among the many observations that led to a better understanding of AKR the most successful were born on spacecraft designed to cross the nighttime auroral region: Viking \citep{hultqvist1990}, Freja \citep{wahlund1994}, and in particular Fast \citep{carlson1998}. They crossed the source region of AKR providing  important plasma and field information. In addition, AKR was observed by  more distant spacecraft starting with Imp 6 and 8 \citep{gurnett1974} and providing information about solid angle and total radiated power. The main results of this rather incomplete list -- we apologize to those experimenters we, for brevity, do not mention here -- were that AKR is emitted very sporadically and by no means at fixed frequency or location in space. It moreover seems to consist not of one source but of at least several (if not a large number of elementary, see below) sources whose radiation mixes when observed from distance to form the common integrated emission band of AKR. All these properties are rather difficult to be explained by one single and presumably simple radiation mechanism located in one single source even when spatially distributed. Nevertheless, the picture which ultimately emerged and has become accepted as the canonical emission mechanism for AKR (afterwards mapped also to the other magnetized planets and their magnetospheres \citep{zarka2005,hess2009} as well as to some problems of solar radio emission \citep{zhang2022} and strongly magnetized astrophysical objects)  is the electron-cyclotron-maser instability (ECMI). Here, for our purposes, it suffices to mention just some key observational properties of AKR, before going ahead to a brief discussion of this main accepted radiation mechanism. }

\begin{table}[t]
\caption{\label{list} Selected spacecraft that contributed to AKR observations} \vspace{1.5ex}
\begin{tabular}{|c|c|c|l|}
Spacecraft &  $\sim  r/\mathrm{R}$ & Instruments & Reference \cr
\hline
&&&\cr
Imp 6 & $\sim -30$ & waves & \citep{gurnett1974}\cr 
Voyager 1\& 2 & & waves/plasma & \citep{kaiser1978}\cr
AMPTE IRM & $\pm(15-18)$ & waves/plasma & \citep{labelle1989} \cr
Geotail & $-(10-30)$ & PWI/LEP & \citep{nishida1994,kokubun1994}\cr
&&&\citep{mukai1994,matsumoto1994a,matsumoto1994b} \cr
CLUSTER &$>|\pm10|$ & PWI/plasma & \citep{mutel2003,balogh1997}\cr
&&&\citep{yearby2022}\cr
STEREO & & waves & \citep{panchenko2009} \cr
MMS & $\sim -20$  & PWI/plasma & \citep{burch2016,torbert2016}\cr
 Cassini & $1.2$ to $-6400$ & PWI & \citep{anderson2005,lamy2010} \cr
 Wind & $<|\pm300|$  & waves/plasma & \citep{bougeret1995,lepping1995}\cr
 &&&\citep{wilson2021} \cr 
 &&&\cr
 \hline
 &&&\cr
 DE 1 & $< 5$ & PWI/plasma & \citep{mellott1986,hoffman1981}\cr
 &&&\citep{mellott1984}\cr 
 Hawkeye & $<3$ &plasma &  \citep{green1977,voots1977}\cr
 Fast & $<2$ & PWI/plasma & \citep{carlson1998,carlson1998a,pfaff1998}\cr
 Viking & $<3$ & PWI/plasma & \citep{hultqvist1990,ungstrup1990}\cr
 &&&\citep{bahnsen1989,roux1993}\cr
 Freja  & $<2$ & PWI & \citep{wahlund1994,louarn1994}\cr
 S3-3 & $<2.5$ & PWI/plasma & \citep{mozer1977,temerin1981}\cr
 &&&\citep{temerin1984} \cr
 \end{tabular}\\ [1ex] 
 \small{PWI = \emph{Plasma Wave Instrumentation}, LEP = \emph{Low Energy Particles}\\
 waves = \emph{plasma waves}, plasma = \emph{general plasma measurements}}
\end{table}%

{Spacecraft observations of AKR can be divided into those crossing the presumable AKR source region and those observing AKR remotely. Table \ref{list} is organized just in this manner (the second column gives the geocentric distance). Moreover, in the Table the spacecraft passing close to earth were at high geomagnetic inclination crossing the auroral zones many times when precessing around earth. At large distance from the auroral zone AKR propagates in free space just being subject to reflection and refraction if encountering sufficiently dense environments and bent magnetic fields such that the propagation angles in the dispersion come into play. The radiation flux turned out to decrease inversely to the increase in traversed surface area \citep{gurnett1974} suggesting conservation of flux which led to the estimate of the radiated power given above. From those large distances little can be said about the radiation mechanism except its polarization and that the space-time variability correlated well with the state of disturbance of the magnetosphere. The `lightning' of AKR seen from remote indicates that something violent is going on in the auroral zone. This is an important insight from the viewpoint of Space Weather and the global inference it provides. More recently some of the related questions have been considered in a statistical sense in order to infer about the coupling of the solar wind and the magnetosphere \citep{fogg2022,waters2022}. These investigations do not intend to contribute, however, to the mechanism of generation. They are of large value in the macro- and mesoscopic physics of the magnetosphere and should become a continuous observational tool to monitor its state in a global `Space Meteorology' making use of  the available fleet of remotely earth orbiting satellites at all distances, including non-orbiting spacecraft located at large distance like Wind in a solar wind Lagrange point. We note that in addition to the spacecraft in Tab. \ref{list} some other satellites occasionally reported observations of AKR from remote and closer in (among them were ISIS 1 \citep{benson1979},  POLAR \citep{menietti2000,mogilevsky2008}), ISEE 3 \citep{benson1991,gallagher1979,imhof2000,hashimoto2016,grach2020,green2004}.)} Galileo \citep{menietti1996}, ERG \citep{kolpak2021}, Exos-B \citep{morioka1981}, Interball \citep{parrot2001,mironov2006,mogilevsky2005,mogilevsky2011,moiseenko2013}, 

{From a more fundamental physics point of view spacecraft crossing the auroral region  provided invaluable information about the presumable AKR source region. Not only that they confirmed the generation of AKR in the upper auroral ionosphere/lower magnetosphere mainly on the nightside and  below the cusp on the dayside \citep{frey2019}. They found that AKR is generated as a result of the interaction of the disturbed magnetospheric plasma component when it interacts with the upper ionosphere. Essentially four spacecraft contributed to clarify the role of the plasma: S3-3, Freja, Viking, and Fast. The first of them, S3-3 demonstrated that the lower magnetospheric auroral convection pattern generates larger scale localized field aligned electric potential drops caused by convective shear and result in dilutions of the upper ionospheric plasma, so-called electrostatic shocks and auroral cavity. These have provisionally been interpreted as Alfv\`en wave structures connecting the magnetosphere to the ionosphere \citep{mozer1977,temerin1981,temerin1984,hilgers1992}. The most convincing assumption is that the magnetospheric source (of Alfv\`en waves and/or downward electron fluxes/field aligned currents) is to be found in reconnection regions evolving in the central-tail plasma-current sheet, mostly during magnetospherically disturbed times though also on the dayside probably originating from below the cusp. Subsequent observations, in particular by Fast, found a wealth of structure in the upper auroral particle distribution in relation to the generation of AKR. Of the large number of their observations and inferences we just note those which are most relevant here for our purposes.}

{Fast confirmed the shear in the convection and the generation of convection boundaries which carry field-aligned electric potentials \citep{carlson1998a}. It found that such cases are stacked in certain order along a poleward track of the spacecraft, dividing the auroral upper ionosphere/lower magnetosphere into a sequence of upward and downward current regions with particular typical particle structure: relatively broad latitudinal regions of downward fluxes of $\gtrsim 10$ keV electrons flanked by narrow highly structured regions of intense upward $\lesssim$ few keV electrons, the former flowing in from the magnetosphere, the latter accelerated upward from the ionosphere by downward electric potentials in the sheared magnetospheric plasma convection. In the downward current region, so-called Debye-scale (along the magnetic field) electric potentials \citep{ergun1998} were detected which are believed to be ion and electron phase-space holes (BGK-modes \citep{davidson1972}), respectively, depending on their polarization, size, potential, and propagation speed. It was later inferred \citep{pottelette2005} that similar structures exist in the upward current region as well there being pure electron holes. Both regions were attributed to Alfv\'en wave flows from and to the magnetosphere. Concerning AKR it seemed that the upward current region acted as its dominant source \citep{strangeway2001} a conclusion apparently supported by the observation of (weak or partially filled) loss-cone distributions in the higher energy electron component there \citep{delory1998} which was in accord with Viking observations \citep{louarn1990,louarn1994}. Those electrons are of magnetospheric origin flowing down into the ionosphere. Being subject to the mirror force and experiencing the field-aligned electric potential drop caused by the convective shear flow, the energetic electron phase space distribution develops a one sided loss-cone. However, the observations show that this loss-cone is rather narrow and at least partially filled which questions its effectiveness. On the other hand, the plasma observations which depend on the spin of the spacecraft, had so far rather low angular as well as energy resolution which may obscure the much more resolved though highly fluctuating structure of the angular and energy distribution. Much higher resolution instrumentation is required.}

{Fast also observed the emission of AKR in high resolution (see  Fig. \ref{fig2} below) which, however, was highly structured in frequency and time, very narrow-band $\Delta\omega\sim 1$ kHz or even less, and in addition sporadic and temporarily variable, obviously being composed of several or even many contributions from different elementary emission sources which to locate was impossible. The narrow bandwidth of the elementary emissions was surprising. At a nominal electron cyclotron frequency of say $\omega_{ce}/2\pi\sim 300$ kHz it amounted to only $\Delta\omega/\omega_{ce}\sim\Delta\omega/\omega\sim10^{-3}$ for emission at the fundamental of the gyrofrequency, suggesting that the elementary emission is very close to the gyrofrequency. Referring to the light crossing time $\Delta t\lesssim 10^{-3}$ s would yield a source extension of $\ell\lesssim 100$ km or even less, because near $\omega_{ce}$ the X mode dispersion relation $\omega_X(k_\perp)\approx \omega_{ce}$ with $ k_\perp c\gg1$ levels out to become about flat (or close to), the phase velocity $\omega_X/k_\perp\ll c$ is strongly reduced, this is not unreasonable.} 

We should add at this point that reviewing all the even very important observations here (like those of the bursty nature of AKR \citep{moiseenko2013,mutel2003,mutel2006}, coupling to the ionosphere \citep{seki2005}, radiation belts \citep{xiao2016,zhao2019}, harmonics \citep{mellott1986} and many others)  is beyond our intention and possibilities. The interested reader is referred to the citations in the listed references.

{This brief account for the properties of AKR and its source region should suffice for our purposes here.  We remind the reader that any radiation can penetrate into a plasma only up to its cut-off frequency which depends on the magnetic field and density. If the X mode is excited in the upper ionosphere close to the electron cyclotron frequency it will not be able to penetrate down to the ground, because the dense ionosphere represents an impenetrable layer at any of the relevant radio frequencies. It was therefore not a small surprise that observations on the ground in Antarctica seemed to suggest that some AKR might have leaked down to the ground \citep{labelle2011}. Thoroughly controlled experiments did indeed confirm this observation \citep{labelle2015,labelle2015a,chang2022} which so far has remained unexplained though some mechanisms for downward transport of radiation based on the ECMI \citep{treumann2012a} have been proposed but have not been proved nor confirmed experimentally yet.}

To close this brief and incomplete overview section on observations, we summarize the main observational results which we consider being of relevance for the purposes of this commentary:
{\begin{itemize}
\item [-] AKR source in the upper auroral ionosphere/lower auroral magnetosphere at several $10^3$ km altitude
\item [-] Its main source location is on the night side and below the dayside cusp
\item [-] It is emitted (mainly) in the free-space X mode beamed perpendicular to the magnetic field 
\item [-] Emission is highly sporadic, variable in time, frequency, and location 
\item [-] Its total maximum power amounts to $\lesssim 1\%$ of a substorm, but values $10^{-2}\%$ are more frequent
\item [-] At large distance from earth its flux decreases inversely to the crossed surface area
\item [-] Source: low-density auroral cavity in a sequence of up- and downward auroral current regions 
\item [-] Either downward electron loss-cone distributions of $\gtrsim 10$ keV electrons
\item [-] or upgoing $\lesssim$ few keV ionospheric electrons are responsible for AKR generation
\item [-] Also involved convection shear, field aligned potentials (electrostatic shocks) and 
\item [-] Debye-scale structures (phase space holes) which are observed in multitude in the source region
\item [-] AKR is structured in at least several drifting bands of $\Delta\omega_X\sim 1$ kHz
\item [-] Sometimes AKR seems to leak down to the ground as bursts of weak radiation
\end{itemize}}

\section{Radiation and AKR} 
Radiation in the radio frequency range is an indispensable tool to remotely sense distant magnetized objects not accessible to direct observation in situ. The magnetosphere is one of the rare objects where in situ observations/measurements can be performed. It thus provides the \emph{{unique} opportunity} of checking theories/hypotheses about the mechanisms of {nonthermal radio} radiation generation. 

Thermal radiation poses no serious problem as long as the coefficients of emission and absorption are known, which is the case in dense media as well as in dense plasma like the interior of the sun. Radiation in this case diffuses across the matter, which is described by the equations of radiation transport, until outside matter in empty space where one applies one of the relevant approximations to Planck's law. In dilute collisionless plasmas like the upper auroral ionosphere and magnetosphere thermal radiation is very weak. Some calculations, theoretical and numerical, suggest that for finite temperature it is not zero \citep{yoon2017} though much too weak and about stationary as to apply to AKR, which clearly is the result of some non-thermal excitation mechanism. The finite weak thermal background, by itself providing a continuous radiative energy loss does however provide a radiation bath from which any nonthermal mechanism can choose to amplify escaping radiation.

The general condition on radiation in that case is that it, after being excited locally, escapes into free space only if it manages to jump into one of the free-space (magneto-ionic) modes: O, X, Z, which have different polarization and properties of propagation \citep{budden1988,melrose1980}. For instance the ordinary (O) wave  $\omega_O>\omega_{lc}$ propagates above the lower frequency cut-off $\omega_{lc} <\omega_{ce}$ defined (with minus sign in the brackets) as
\begin{equation}
\omega_{lc, uc}=\textstyle{\frac{1}{2}}\Big[\Big(\omega_{ce}^2+4\omega_e^2\Big)^{1/2}\mp\omega_{ce}\Big]
\end{equation}
( $\omega_e,\omega_{ce}$ are the respective local plasma and electron cyclotron frequencies) almost parallel to the external (homogeneous) magnetic field (implying that in a curved field it may become oblique). The extraordinary (X) wave $\omega_X>\omega_{uc}$ propagates perpendicular to the external magnetic field but only above the upper cut-off $\omega_{uc}>\omega_{uh}$ where $\omega_{uh}$ is the upper-hybrid frequency, and the Z-mode, propagating as well perpendicular, is confined to the region between $\omega_{uh}<\omega_Z<\omega_{uc}$ which usually forms a rather narrow band only. 

Thus Z radiation (which classically is not a real radiation as it does not escape from its band and has characteristics closely related to an electromagnetic partner of the upper-hybrid wave, its electromagnetic branch) should only occasionally be observed, mostly by chance in situ, which has indeed been claimed sometimes. 

The majority of observations, however, finds from its polarization that AKR is in the X mode and barely in the weaker O-mode polarization. Importantly, the X-mode has a lower frequency branch which is confined to the plasma to frequencies $\omega<\omega_{uh}$ at large perpendicular wave numbers, becomes slow with speed $\omega/k<c$, and cannot leave to free space unless crossing the gap to the free-space branch of X.

Any AKR mechanism thus has to explain how the X-mode can be excited above $\omega_{uc}$ either directly in a plasma or, if excited by some process below $\omega_{uh}$, how it can jump across the frequency gap $\omega_{uh}<\omega<\omega_{uc}$ in order to reach the upper X-mode branch. 

Generation of radiation in the free-space modes can be done in various ways. The simplest is by a direct wave-wave interaction process of three (or more) waves under energy $\omega=\omega_1\pm\omega_2$ and momentum $\mathbf{k}=\mathbf{k}_1\pm\mathbf{k_2}$ conservation, a perturbation-theoretical approach in so-called weak plasma turbulence theory \citep{davidson1972,melrose1980,sagdeev1969,treumann1997}, where the free space mode frequency $\omega$ must be above the upper cut-off $\omega_{uc}$, and the colliding frequencies and wave numbers must satisfy the relevant dispersion relations of plasma waves. This works pretty well in a dense plasma like the solar wind for two colliding Langmuir waves of opposite direction excited for instance by a gentle-electron-beam instability. It explains the generation of so-called solar Type-III and spike bursts \citep{lin1981,farrell1985,melrose1986,melrose1994} and in similar way explains radiation emanating from a shock like the earth's bow shock wave \citep{eastwood2005,balogh2013} where this kind of radiation has {almost continuously  been observed at the foreshock boundary of earth's bow shock, forming the paradigm of weak nonthermal radio-radiation from plasmas, in this case excited by reflected or accelerated electron beams from the shock. }

In AKR this mechanism does not work, however, because the plasma is highly underdense with plasma frequency $\omega_e\ll\omega_{ce}$ too low, even though intense energetic auroral electron beams are present. It could appear at very high harmonics $\omega=n\omega_e$ ($n\gg1$) only, where it becomes very weak. It thus can be excluded as mechanism of generation of AKR, though other electron-beam excited waves have occasionally been made responsible, yielding very low radiation efficiencies. in fact they barely generate radiation but low frequency waves, electrostatic and electromagnetic which are not primarily of interest in AKR but may secondarily contribute to it in various {ways} by modulation, damping, structuring the plasma in density and accelerating/retarding the beam electrons with the effect that their energy is dissipated. In each case, however, the conservation laws have to be accounted for which for most plasma waves leads to severe difficulties in satisfying energy (frequency) and momentum (wave number) conservation both in the elementary wave-wave interaction or in integration over the spectrum, when accounting for the complicated dispersion relations. Overcoming this difficult requires taking into account the propagation properties in inhomogeneous media \citep{zarka1986}. All these processes are second or higher order and thus barely efficient enough for explaining the comparably high AKR intensity.

\begin{figure*}[t!]
\centerline{\includegraphics[width=0.5\textwidth,clip=]{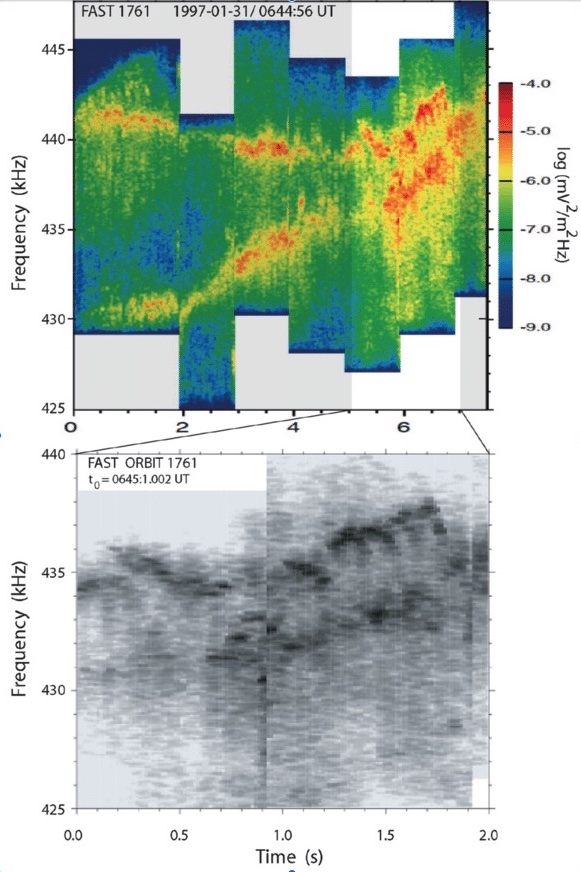}}
\caption{High time and frequency resolution FAST observations of the fine structure in the spectrum of AKR during passage of the auroral region (adapted from \citep{treumann2012}).} \label{fig2}
\end{figure*}

\section{Electron-cyclotron maser -- the canonical paradigm}
Electrodynamics tells that emission of electromagnetic radiation from particles is a third-order particle process. This is a well-known fact \citep[see e.g.][]{rybicki1979}. Any direct emission from plasma is generally weak. In order to become intense it requires to be coherent. This is the basic idea in maser and laser theory where electrons are lifted to higher than thermal energy levels by some process of energy pumping. Within the life-time of the higher level the electrons then jump down to release their energy coherently in radiation, as was suggested by extraordinarily successfully re-interpreting Einsteins A-B coefficients of absorption and emission. It was first speculated to apply \citep{twiss1958} as well to plasmas. In an extended collision-dominated medium, pumping by external radiation leads to inversion of absorption as happens in some astrophysical configurations (dense molecular clouds illuminated by uv-radiation, for instance). 

Any kind of such pumping does not exist in the dilute collisionless plasma even of the lower auroral magnetosphere (unless one could attribute it to upward lightnings resulting from atmospheric discharge as these transport sufficient energy, a mechanism which has never been considered yet in the generation of AKR or any other magnetospheric effect). The source of AKR is thus to be sought in special forms of the electron particle distribution leading to coherent emission. 

The mechanism that ultimately has been envisaged is the electron-cyclotron maser radiation \citep[for reviews see][]{hirshfield1977,sprangle1977,chu2004,treumann2006}. It had been proposed first under the above mentioned assumption of a large temperature anisotropy in the auroral electrons \citep{melrose1976} as an equivalent to lifting the electrons collectively into a higher level providing sufficient free energy to induce a substantially intense quasi-coherent radiation. Coherence in this case means that the instability is resonant and the bandwidth of emission small such that a limited number of electrons in the particle distribution are locally simultaneously involved. Note that this is just a simple explanation in passing. In fact the radiation is simply a linear instability of large growth rate $\Gamma(\omega)$ with involvement of a certain number of resonant electrons. (Large here means that, for reaching high wave intensity of the quickly escaping radiation with $\omega/k\sim c$, the growth rate must be of the order of the wave frequency $\Gamma\gtrsim\omega$, a condition which is very difficult to satisfy.) However the temperature anisotropy suggested must be huge in terms of the ratio of perpendicular to parallel of the order $T_\perp/T_\|\gtrsim10^3$ or more which (due to the thermal spread of the distribution) is not even satisfied anywhere at the mirror points. This seemed to make this kind of direct excitation of AKR obsolete. 

The resolution came with \citep{wu1979} where it was realised from the work in \citep{hirshfield1977,sprangle1977} that, even for the low energies in the auroral electron component, the relativistic correction must be taken into account. This correction is contained in the relativistic electron-cyclotron frequency $\omega_{cer}=\omega_{ce}/\gamma$ with $\gamma=\sqrt{1+p^2/m^2c^2}$ the relativistic energy factor, and $\mathbf{p}=(p_\|,\mathbf{p}_\perp)$ the relativistic momentum of the electrons.  Since this enters the resonant denominator 
\begin{equation}\label{eq-res}
\omega-\omega_{ce}/\gamma-k_\|v_\|\approx0
\end{equation}
with complex frequency $\omega=\omega_X+ i\Gamma(\omega_X)$ for linear instability $\Gamma>0$ a velocity or momentum dependence appears in the resonance even for purely  perpendicular propagation $k_\|=0$ of the resonant wave. The effect is that the relativistic resonance $\omega_X(k_\|=0,k_\perp)\equiv\omega_X(v_\perp,v_\|)=\omega_{ce}/\gamma$ in this perpendicular X-mode propagation case becomes an ellipse in velocity space which is to be integrated along over all particles contributing to it. The frequency is of course the dispersion relation of X-mode, while the kinetic resonant dispersion relation determines the growth or damping rate of the wave. Hence the problem of obtaining an exponentially amplified X-mode depends on cleverly modelling the auroral electron distribution function in such a way that for positive perpendicular velocity space gradient $\partial f_e(v_\|,v_\perp)/\partial{v_\perp}>0$ of the auroral electron distribution function $f_e(v_\|,v_\perp)$ along the resonance ellipse as many electrons as possible with parallel speed $v_\|$ contribute positively to the growth rate $\Gamma(\omega_X)>0$ causing exponential growth of the X-mode amplitude at resonant frequency $\omega=\omega_X$. Note, however, that $\Gamma\sim\omega$ must be large, of the order of the wave frequency, as was mentioned above! Otherwise the wave escapes from the source with very little amplification. Fortunately this restriction is substantially relaxed by the intrinsic radiation mechanisms, as we will shortly demonstrate.

Positivity of the perpendicular gradient of the distribution in velocity space along the resonance ellipse is required for wave excitation. It perfectly corresponds to an inverted energy state in an average statistical sense where particles at low energy are absent (or at least strongly reduced in density) below the resonance line in velocity space, and the resonance ellipse is placed at the largest perpendicular phase space gradient. Of course, the entire region of positive perpendicular velocity space gradient contributes. Its velocity space width $\Delta v_\perp$ determines the bandwidth $\Delta\omega(\omega_X)$ of emission at constant frequency.

Since the X-mode predominantly propagates in perpendicular direction to the ambient geomagnetic (magnetospheric) field the resonance ellipse contains merely the perpendicular wave number  $k_\perp$ through the dependence of the X-mode frequency. $k_\|$ comes into play only for large $ v_\| < c$ where in the non-relativistic auroral plasma the particle number is exponentially small. 

{Under these conditions it turns out that the most efficient particle distribution causing AKR in the X-mode could be provided by kind of a loss-cone distribution \citep[as observed,][]{delory1998} in the resonant particle component  \citep{pritchett1984} with possibly a wide loss-cone and a very diluted absorbing electron background. Clearly this comes close to the nonrelativistic proposal \citep{melrose1976} but firstly corrects it to the relevant relativistic case \citep{pritchett1986,pritchett2002} and, secondly, relaxes it substantially to more realistic electron distributions. However, all observations like \citep{delory1998} find only very narrow and even partially filled loss cones, unfortunately suggesting that few electrons are involved into the cyclotron maser instability. We may note at this place that it is quite natural that the loss-cone is narrow in the auroral region because loss-cone distributions are subject to the excitation of intense low frequency short wavelength VLF waves in the whistler band \citep{labelle2002} which quasilinearly depletes the loss-cone not leaving much velocity space gradient for the ECMI and AKR. This observation alone put{s} severe doubts on the real importance of the loss-cone distribution for the maser and AKR whereas the role it plays in VLF is well established for long already. In fact considering the resonance curve in velocity space which is an ellipse shifted in parallel direction one would rather believe that a \emph{shifted hollow} electron phase space distribution would fit considerably better than the loss cone. The shift along the field would correspond to the relativistic parallel momentum of a hollow beam moving along the magnetic field with the gap at low velocities produced due to the combined action of an electric potential and the magnetic mirror force.}

{Hence, except for the observed narrow and partially filled loss cones, which are well understood when accounting for the strong VLF \citep[cf. again][]{labelle2002}, all this sounds satisfactory what concerns the maser and AKR: we have a weakly relativistic mechanism already directly exciting (linearly amplifying) the X-mode correctly into perpendicular direction causing its exponential growth if only the right electron distribution (either with empty loss-cone or otherwise) is present. }Then all modes at perpendicular wavenumbers $k_\perp$ whose resonance ellipse crosses the positive perpendicular velocity space gradient $\partial f_e/\partial v_\perp>0$ will become excited. The resonance width in frequency $\Delta\omega$ and wavenumber are then determined by the number of resonance ellipses which fit into the positive gradient interval $\Delta v_\perp$. Therefore the emission band will probably be narrow and intense for steep velocity space gradients, while becoming weak and broader for flatter gradients. The headaches caused by the unsatisfactory loss-cone point either to a so far unsatisfactory resolution of the angular dependence of the electron distribution, a problem of measurement, and at the same time to a problem with theory when being fixed to the loss-cone distribution. In fact the latter has been used for convenience and to agree with observations in the upward current region. This is not at all required in the downward current region where the loss-cone distribution is absent. 

In wavenumber space this becomes more subtle because the X-mode dispersion relation must be considered as well. Observationally this is not as relevant because measurement of the wavenumber spectrum of AKR has not yet been performed and is technically very difficult. What however is suggested by this kind of model theory is that, at fixed wavenumber $k_\perp$,  the frequency spectrum should approximately map the local width of the perpendicular velocity gradient, whether or not this would be an important information. The total width of the observed broadband AKR should then be somehow related to the spatial extension of the resonant electron distribution (its positive perpendicular velocity space gradient) giving rise to emission, i.e. the spatial size of the radiation source.

In order to know the latter one needs to know the central radiation frequency. This however is a very sensitive point because in considering the resonance conditions (\ref{eq-res}) one steps into a trap which is not as easy to overcome. Consider the ideal X mode case $k_\|\approx 0$. Since $\gamma  > 1$ the instability and consequently the emission occurs at positive (wave energy) frequency 
\begin{equation}
\omega\approx\omega_{ce}/\gamma < \omega_{ce} 
\end{equation}
implying that it is located at the \emph{lower} X-mode branch which principally cannot escape from the source, i.e. from the plasma into free space. Even in an underdense plasma like that of the auroral cavity this becomes a fatal restriction, and one has to call for other effects to enable escape. 

On the other hand, just this restriction seems to relax the problem of the above mentioned condition on the growth rate, which required that $\Gamma\sim\omega$ in order to obtain sufficiently high amplification rates. In fact, growth of the wave on the lower X-mode branch (indexed $l$) close to $\omega_X^l\lesssim\omega_{ce}$ implies that the wavenumber of the excitation $k_\perp$ is very large, because the lower X-branch dispersion curve flattens out near $\omega_X^l\sim\omega_{ce}$. This means that the phase velocity of the  X-mode at instability on the lower branch becomes very small, in fact small enough for the wave to stay so long in the source region that, even for  $\Gamma\ll\omega\sim\omega_{ce}$, the growth can become substantial by allowing for many e-foldings. This is a fortunate fact. One thus excites a highly intense X-mode near $\omega_X^l\lesssim\omega_{ce}$ on the lower branch, but this wave has very short wavelength and, in addition, cannot escape unless other processes make escape possible. 

\begin{figure*}[t!]
\centerline{\includegraphics[width=0.5\textwidth,clip=]{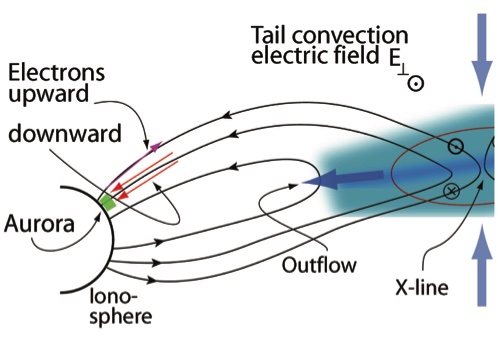}}
\caption{Sketch of the auroral magnetosphere with upward and downward electron fluxes (red arrows) originating from the tail crossing the presumable source region of AKR, which is above the aurora (green spot) and some variable fraction of the electron arrows long around roughly 1000 km to 3000 km altitude (above earth's surface). This region is connected to the tail plasma sheet via the geomagnetic field with downward accelerated electron fluxes resulting from reconnection. The downward currents are transported by upward cool ionospheric electrons of higher density (adapted from \citep{treumann2017}).} \label{fig3}
\end{figure*}

The simplest escape from the trap is assuming emission close to a higher harmonic of the electron cyclotron frequency $\omega\lesssim n\omega_{ce}$, with $n>1$, a not unreasonable assumption as it still remains uncertain whether AKR is excited at the fundamental or any higher harmonic. (Note that higher harmonics due to this simple mechanism have been favoured in \citep{wu1983}.) The caveat here is that the growth rate of the harmonics steeply decreases with harmonic number $n>1$. Nevertheless this kind of mechanism is clearly viable. It is not unreasonable to assume that indeed just low harmonics are excited, not the fundamental X-mode. 

Recent numerical calculations of the thermal level of perpendicular electro-magnetic Bernstein modes in magnetized plasma \citep{yoon2017} suggest that the thermal noise at the magnetic cyclotron harmonics structures harmonically like $\omega_{th}=n\omega_{ce}$. It starts already at $n=0$, for every harmonic number filling almost half the harmonic band above and below $n$ with increasing wave number $k_\perp$  and is of finite amplitude (which in itself is a most interesting and important observation in as far, as none of the instability theories proposes the existence of instability in the first harmonic band below $n=1$. Unexplained half-electron-cyclotron harmonically structured noise in the lowest band $n=0$ has indeed been observed by the wideband receiver on the AMPTE IRM spacecraft \citep{labelle1990}). This thermal fluctuation provides sufficient thermal background which by some instability (in the present case the electron-cyclotron maser instability) can become amplified picking up its required wave number range. The electron-cyclotron maser instability may readily take advantage of the presence of the thermal fluctuation background in its appropriate wave number range at lower harmonics $n\geq1$ of the electron cyclotron frequency and let it grow quickly to observable amplitudes. This radiation will be weak because the wave disappears quickly into space.

We should, however, note that the problem of escape could also be resolved if one assumes a three wave process acting on the lower branch near resonant amplification. If two short wavelength unstable lower branch X-modes (indexed 1,2) which have been excited by the electron-cyclotron-maser instability merge (in the above described way, grown from background fluctuations below $n=1$ to large amplitudes) and being of opposite direction, such that their large wave numbers cancel, then the process 
\begin{eqnarray}
\omega_X^{u}&\approx &\omega_{1}^{l}+\omega_{2}^l\cr
k_\perp^u&\approx& k_{\perp 1}^l-k_{\perp 2}^l\ll k_{\perp 1,2}^l
\end{eqnarray}
would -- if feasible -- directly generate an escaping long wavelength free-space X-mode wave on the upper branch (indexed $u$). This process has, to our knowledge at least so far, not yet been investigated but might be promising in explaining escaping X-mode radiation. It naturally causes long perpendicular wavelength harmonic emission at $\omega^u_X\approx 2\omega_{ce}$ in the free-space X-mode. {Though the three-wave process is second order, it has some advantages over linear excitation as we are going to show below.}

{What is important is that, for escape, it does not require any confinement or multiple scattering of the unstable wave. Confinement below $\omega_X^l\lesssim\omega_{uh}$ is trivially given; it requires simply that the two oppositely directed waves are generated in the resonant electromagnetic electron-cyclotron instability. Since at $k_\|=0$ the resonance is independent on wave number, the resonance condition only involves the lower branch dispersion relation $\omega_X^l(k_\perp^l)$ which allows for amplification of waves in any perpendicular direction $\pm k_\perp^l$. In addition it profits from the slow phase and group speeds of the excited waves. These unstable waves are nearly immobile and can thus be excited to large intensities before moving away. This  favours their interaction in the above three-wave process. In addition, reaching high intensity they would affect the environment over scales of their perpendicular wavelength thereby causing nonlinear side effects like the excitation of ion-acoustic waves and/or solitons before interacting in the three-wave process and escaping  into free space at harmonic number $n=2$. It is obvious that a large number of processes like these have not yet been investigated. They all might be of non-negligible relevance in magnetospheric physics as well as paradigmatically in other remote and barely accessible systems in space and astrophysics including the large magnetized planets, exoplanets, the sun, and galactic as extra-galactic non-thermal radio sources.}

Another final possibility for escape is that the parallel wavenumber remains finite and parallel propagation is parallel to the parallel resonant electron velocity  such that $k_\|v_\|>0$ which from 
\begin{equation}   
\omega\approx\omega_{ce}/\gamma + |k_\|v_\|| > \omega_{ce}
\end{equation}
for large parallel speeds $v_\|$ may compensate partially for the smallness of $k_\|$ and the relativistic reduction of the electron cyclotron frequency $\omega_{ce}$, which for slightly oblique propagation of the X-R-mode then allows bridging the electromagnetic stop band of the X mode. The latter condition determines, for given small $k_\|$, the parallel velocity and thus affects the form of the resonance curve. Still, $k_\perp$ is large on the lower branch for resonance below $\omega_{ce}$ which implies that the instability is very short wavelength and very close to but below  $\omega_{ce}$. Bridging the gap requires a large correction to the frequency in order to reach the free-space upper X-mode branch which means large parallel velocities of the resonant electrons on the resonance curve!  Accounting for the dispersion relations of the two X-R-modes below and above the gap imposes a severe restriction on this process, however, which we do not discuss here as it requires some lengthy algebra. Only in exotic cases this mechanism might work. It in any case results in highly variable and non-harmonic free-space X-mode radiation which is unrelated to the local electron cyclotron frequency or any of its harmonics $n\omega_{ce}$. This may make this mechanism interesting in other applications as for instance the {``mysterious''  so-called  Fast Radio Bursts (FRB, for their brief description see below) in astrophysics \citep{petroff2022} }. 

It is interesting that this kind of bridging somehow corresponds to a classical tunnelling across the stop-band gap. Also interesting is that the parallel displacement of the radiation along the magnetic field in the auroral zone where it is believed that the radiation is generated by downward $>$ 10 keV auroral electrons (see Fig. \ref{fig3} for a sketch of the macro-geometry and structure of the auroral source region) the radiation would have to move into a stronger magnetic field which makes it increasingly more difficult to satisfy the condition of escape. On the other hand, if in the downward current region upward  lower energy electrons (few keV at most, questioning the relativistic assumption) and not obeying a loss-cone distribution, serve as the agent which excites the instability then the radiation has it easier to bridge the gap in outward propagation. Thus this latter case seems to be more appropriate as an explanation of AKR relating its generation and escape to the downward rather than upward current region in the auroral zone. Unfortunately the usual belief is that AKR is generated in the upward current region. This has already been questioned  \citep{pottelette2005,treumann2012} though for different reasons.

In addition to these two physically clean ideas quite a large number of less clean proposals have been put forward. Most of them use arguments based on the inhomogeneity of the plasma assuming that the radiation  after emission moves into regions of lower magnetic field strengths until finding itself on the free-space X-mode branch. This can be achieved for perpendicular radiation by poleward radiation to higher latitudes tangential to the earth. It can also be achieved by multiple scattering of the lower-branch wave into a region farther out from earth. In such multiple scatterings in a curved magnetic field and outwardly opening auroral density cavity the mode necessarily becomes oblique. Still, in this process the caveat remains unresolved that the wavenumber of the unstable lower X-mode near $\omega_{ce}$ is very large, and thus the free space mode ultimately reached must be fairly high frequency, nearly unrelated to the original electron cyclotron frequency of excitation.
  
\section{AKR fine structure}
Measurements of electron fluxes in situ the auroral radiation source region, the upward and downward current regions, in particular by the FAST spacecraft \citep{carlson1998b}, revealed that AKR possesses a distinct highly dynamical internal fine structure that is strongly related to the auroral particle fluxes. This observation basically destructed any stationary large-scale models of generation of AKR. It did not vindicate the assumption that the loss-cone might contribute to the generation of AKR but demonstrated that more subtle properties of the particle fluxes play an important role in the AKR source \citep{ergun1998}. The question for these sources is indeed left open until today.

Though it is clear and has been proven by all observations, in particular those from FAST, that the auroral particle fluxes along the auroral magnetic field provide the source of energy for AKR as proposed in \citep{gurnett1974} and confirmed in any of the many follow-up observations which we do not list here, it is also quite clear that the loss-cone distribution barely can account for any of the  AKR fine-structures, time variability and narrow bandwidth as well as the dynamics of the emission. 

An example is given in Figure \ref{fig2} which shows two narrow-band AKR emission events approaching each other. The black-and-white lower part of the figure is a brief sequence of the occasionally available highest time resolution measurement $\Delta t\sim 0.03$ s. It indicates that the elementary emissions are close to this resolution with $\Delta t\sim 0.05$ s. The bandwidth of emission at largest intensity is rather small $\Delta\omega<1$ kHz which must be attributed to the properties of the exciter, from the resonance condition the perpendicular velocity space gradient, thus the subtleties in the electron phase space distribution. The upper panel suggests that the total bandwidth (green) of the AKR emission in this event, though  variable in time and frequency over the nine seconds of observation, amounts to roughly 25 kHz (presumably at the first harmonic $n=1$ in this case as inferred from other properties we do not discuss here). It consists of the weak green background and two intense narrow emission lines $\Delta\omega<2$ kHz which in time approach each other while do not merge. What happens in this case is unclear; below we list a few explanations. 

To make it short, the loss cone boundary of the auroral electrons is, of course, highly variable in time, velocity space gradient, and location.  It will fluctuate in all these parameters. Whether any such fluctuation could contribute sufficiently to come up for the excitation of the observed intensity is not known and difficult to check. It depends on the form of the resonance line, the number of resonant electrons involved and the time available for amplification. It also depends on the accuracy of the electron measurements \citep{delory1998} and resolution in direction, energy and time which is technically restricted. However, possibly the farther above mentioned hollow beam is a better suited model of the electron distribution. It has so far not been identified in the observations but probably requires better angular and energy resolution. 

The local emission frequency corresponds to an oscillation time of $\Delta t\sim 10^{-6}$ s which allows for $10^3-10^4$ wave oscillations at any of the local (dark red, black) emission maxima. The amplification over background  amounts to roughly four orders of magnitude in intensity (as read from the colour bar). Linear growth then suggests a growth rate of roughly $\Gamma_X\sim 2\times10^{-4}\omega_X$ at $\omega_X/2\pi\sim 440$ kHz, a large but not unreasonable growth rate requiring a rather strong emission mechanism/steep velocity space gradient and large local number of resonant electrons. Whether fluctuations in the loss cone could yield such values remains a tantalising  open question. Still the question remains for the strange dynamics of the local emissions in the narrow emission bands. 

{The local radiation events are bent in time. They, in this event, move in time up in frequency, reach maximum and subsequently move down in frequency again. Can any displacement in the loss-cone along the field explain the systematic upward-downward dynamics in the spectrum? Or does it result rather from propagation of the emitted radiation i.e. as the consequence of a simple Doppler effect? That any fast changes in the loss-cone distribution could explain and provide it, seems improbable. Filling the loss-cone proceeds on a long quasilinear time-scale \citep{labelle2002} and should cause a drop out of radiation, resulting in an emission gap on the loss-cone filling-time scale. The systematic symmetric pattern rather resembles the action of Doppler passage toward observer (increasing frequency) and subsequent turn-away (receding frequency). Emission is strongest at closes approach. This might be the easiest interpretation containing information about the relative speed of the radiation source and its spatial extension (which in itself is very interesting to infer though we do not go into any detail, here). But what then causes the regularly pulsed emissions with $\Delta t\sim 0.1$ s pulse length equal to a modulation of frequency $\omega/2\pi\sim 10$ Hz somewhere in the micropulsation range \citep{hanasz2008}, possibly suggesting a typical time-scale of auroral electron injections causing loss-cone structures? Maybe. Altogether, the entire background seems to consist of radiation from very many such structures (elementary radiation events).}

If this Doppler-interpretation is true then, at least in the radiation event shown in Fig. \ref{fig2}, the AKR background is produced by a wealth of such moderately remote elementary radiators of which some of them approach the spacecraft. These must be organized in a narrow stream that passes very close to the spacecraft. This observational fact is independent on whether or not the loss-cone is responsible for the generation of the radiation.  Any other electron distribution capable of generating elementary emission events will do.  

If the loss-cone model does not satisfactorily work to generate those elementary emission events (which may be the case but has neither been proven nor dismissed yet),  which other mechanisms can be envisaged then? Several always rather complicated or exotic ideas have been proposed \citep{strangeway2001,pottelette2001,treumann2017} in particular with the obviously proven observation of so-called Debye-scale structures in the AKR source region, solitons, small-scale double layers, or more precisely electron (and ion) holes \citep{ergun1998}. These are complicated non-neutral electron (ion) structures possessing localized electric potential drops/fields (sometimes called electrostatic shocks) consisting of potential-trapped low energy thermal diffuse dilute electrons in the holes and appear to be sources of higher-energy nearly mono-energetic non-trapped electron beams accelerated along the magnetic field. They usually belong to so-called ``horse-shoe distributions'' which are an electric-field-modified version of ring-distributions in phase space, generated in the concerted interaction of electric potentials along the magnetic field and the mirror force in a converging magnetic field like that in the auroral magnetosphere and upper auroral ionosphere. Moderately steep perpendicular velocity space gradients are naturally produced in this case, and there are very many of those holes and their corresponding horse-shoe distributions in the auroral volume, predominantly  not in the upward but the downward current region. This led to the suggestion of some  models of generation of AKR by those local structures with several rather appealing properties \citep{treumann2012}. 

However, AKR has kilometre wavelength while those Debye-structures are at most of $\Delta z\sim$ few $\lambda_D \ll 1$ km wide \citep{ergun1998,carlson1998b}. Naively it would be hard to believe that one single hole could  excite waves of that lengths. But electron holes are Debye structures only along the magnetic field but can be quite extended in the perpendicular direction with radius of the order of the electron thermal gyroradius much in excess of $\lambda_D\ll \rho_{ce}\sim \sqrt{T_{e}/m_e\omega^2_{ce}}$. Thus their perpendicular scale $\ell_\perp\gg\lambda_D$  may approximately fit the  perpendicular AKR X-mode wavelength on the lower branch near the local electron cyclotron frequency $\omega_{ce}$. This is particularly true for the cyclotron maser instability which excites waves of frequency $\omega_X^l\lesssim\omega_{ce}$ which have very short wavelength and large $k_\perp^l$ (as has been discussed above) which would fit the perpendicular extension of a hole. One thus expects that $k_\perp^l\ell_\perp\sim1$ for those electron holes is not an unreasonable assumption. If true, electron holes could well serve as exciters on the lower X-mode branch, each belonging to one elementary source. Since these modes have low frequency, they are trapped, may grow to larger amplitude and undergo wave-wave interaction until releasing a second harmonic long wave length free-space X mode according to the above mechanism. The escaping radiation clearly maps the local magnetic field at source location. It is moreover also clear that for a long chain of holes (elementary sources) along a magnetic flux tube would map the entire variation of the magnetic field into radiation. If the chain is the result of a field-aligned current or electron flux pulse then the emission would appear as a broadband  pulse mapping the entire magnetic field through the chain of {sources} into radiation and lasting just for the life time of the pulse whose dispersion in time would map the relativistic propagation speed of the pulse.

Moreover, if  one electron hole would still be insufficient to excite one of those short wavelength lower-branch X-modes, combination of holes with their locally strongly deformed velocity-space distribution and horse-shoes may then provide a viable source of generating elementary radiation events. Since the number of electron holes in the auroral source region chained along the magnetic flux tube close together and perpendicular to the magnetic field in the auroral tubes of electron precipitation (in the upward) or upward acceleration (in the downward current region) is huge, the concerted cooperation of many of them should add up to the observed comparably high (coherent) radiation intensity. Indeed, in the auroral magnetosphere the magnetic field varies only weakly along one magnetic flux tube over the scales of one electron holes. A fairly large number of holes forming a chain along the field experiences about the same magnetic field strength and thus feels the same resonance. Within the bandwidth provided by the perpendicular electron distribution it may excite the same lower branch wave.

Given that any of those mechanisms work, there remain two open questions. The first concerns the nature of the narrow emission lines in Fig. \ref{fig2}. It probably belongs to a large group of related elementary sources that pass close to the spacecraft together contributing to the intense narrow band radiation. The motion of this intense band across the spectrum is thus decoupled from the Doppler effect on the radiation. Since emission should be close to the first harmonic of the cyclotron frequency, the upper emission line in Fig. \ref{fig2} is at approximately constant location in space in the stronger magnetic field at lower altitude than the lower emission line which is at weaker field and thus initially slightly farther out in the magnetosphere. It, however, sinks within the next seconds down to approach the upper line, and when hitting it, the two do not merge but move together downward into a slightly stronger magnetic fields. One may speculate in many different directions why they approach each other, whether or not in the same flux tube, and why they do not merge but move further in tandem. Though these questions are very interesting, pursuing them further here would go too far in the context of the present note.

\section{Possible astrophysical relevance}
{The former discussion glanced the application of the ECMI to the magnetosphere as the primary laboratory of any space physics. It may have become clear that AKR is by far not understood yet neither from the point of view of observation/measurement nor theoretically. It has however crystallized that with rather high probability the ECMI is the correct mechanism for the excitation of intense radio emission with properties observationally detected in AKR even though many questions still remain open. The magnetosphere appears as a power house of this kind of radio radiation and, because of its accessibility, offers the opportunity of its in depth study in view of application to inaccessible remote object in astrophysics. Its broad field of application should open up in astrophysics when dealing with non-thermal radio emission from magnetized objects. 
Such objects must be magnetized, surrounded by a diluted plasma which allows for the generation of field-aligned currents and field aligned particle fluxes while itself performing some kind of convective motion.} 

{Conditions of this kind are natural for the larger magnetized planets in the solar system, they are also expected for a certain family of extra-solar planets. 
Other candidates are flaring stars like the sun where highly dynamical regions of closed magnetic loops reach far out into the atmosphere and generate field aligned currents and flows. Actually, the sun was one of the first objects where the ECMI has been applied though just with rather moderate success so far (for reference see \citep{melrose1986,melrose1994}) because of the lacking knowledge about the particle distribution and insufficient spatial resolution. One may expect that future higher-resolution observations will improve over that. }

{More generally, any accreting heavily magnetized astrophysical object should become an ideal candidate for applying the ECMI to radiation in the appropriate frequency range near the local electron cyclotron frequency. Any such non-thermal emission has so far been attributed to synchrotron emission, in very strong magnetic fields to its quantization in lowest Landau levels. In spite of quantum effects,  convection in the accretion disk of a rotating heavy central object will, similar to the magnetosphere, twist the magnetic field in the disk, cause currents to flow, ignite reconnection, and inject field aligned currents towards the rotating object. Since the accreting inflow comes violently to rest near the object surface, current closure evolves, probably with similar effects like in the auroral magnetosphere: plasma dilution, field aligned electric potential drops, particle acceleration, acting mirror forces, and with high probability excitation of the ECMI as a moderate release mechanism of energy stored in current, convection, and particles as radiation into free space, long before and different from bremsstrahlung which is violently released from the bulk matter when hitting the surface. Processes of this kind have barely been investigated yet, at least to our knowledge even though highly spatially structured radio emission from such objects is monitored. Of course, temporal resolution in astrophysics poses problems different from observing radiation in the environment of earth. }

{ECMI as the paradigm of intense sporadic and direct quasi-coherent nonthermal radio emission process is very different from incoherent synchrotron radiation. While the latter applies to extended accumulations of high energy magnetized plasmas mostly of large scale, the ECMI is highly variable in time and thus more difficult to detect/observe. However increase of time resolution in radioastronomy with instruments like LOFAR and ALMA and the VLBI cooperation in the observable radio frequency range promises many new discoveries in the future. One example is the already mentioned cosmic Fast Radio Bursts (FRB). They were discovered in 2007 \citep{lorimer2007} and are ``mysterious'' in several respects as they represent millisecond short radio pulses apparently distributed all over the sky and thus suggesting that they are not local but possibly cosmological associated with galaxies like M81 where one FRB has possibly been identified with a globular cluster, or with galactic magnetars. A collection of the contemporary observational knowledge about them including a number of model explanations can be found in \citep{petroff2022}. These are broadband radio flashes in the say $0.1-8$ GHz frequency  (m to cm wavelengths) range of weak dispersion, sometimes aperiodically, sometimes periodically repeating. Models based on shock waves or reconnection have been proposed. It also seems that they are related to HXRBs (hard X ray bursts) which suggests that electron beams and/or currents may be responsible for them. This is suggestive of similar processes as those in the auroral magnetosphere and leading to either shock or simple magnetospheric models assuming Alfv\'en waves/disturbances propagating outward from the polar caps of neutron stars. Possibly, in the latter case the ECMI could provide another explanatory mechanim. In any case, whether applicable there or not, the ECMI for weakly or strongly relativistic cases is there and can be taken as an additional astrophysical tool diagnosing certain regions in the universe. Its enormous advantage is that it can be investigated \emph{in situ} the magnetosphere in all its properties including its connection to the substorm, i.e. the reconnection processes in the tail or at the magnetopause, thereby also contributing to their understanding.}

\section{Final comments}
A number of still remaining badly understood facts related to ARK  in ECMI-theory, the now mostly accepted paradigm of generation of AKR, have been noted above. They mostly result from the discrepancies between high-resolution (temporal, spatial, spectral) observations including waves/radiation and the presumably responsible particle distribution, the loss-cone distribution. We pointed on the latter as a weak point. ECMI theory does not necessarily require a loss-cone distribution which is prevalent in the upward current region but absent in the downward current region. The challenge on observations is to pin down which of those regions and under what circumstances is the real source of AKR. Maybe the fixation to the loss-cone is not the best choice. Theory must also come up with alternatives. The theory is quite transparent, however comparison with the observational reality still leaves many questions open. This is a tantalizing situation which, however, can be relaxed since in principle the auroral magnetosphere is easily accessible as experimental ground, and observations can easily be accompanied by theory. ECMI theory already has reached a high explanatory level. However it should be extended to distributions of the kind observed in the downward current region, to come closer to reality.

 Moreover, extensions of the theory, in addition to accounting for all kinds of realistic electron distributions, are needed accounting for the effects of  electric fields on scales as encountered in the localized shear motion of the plasma typical for the boundaries between upward and downward current regions in the auroral magnetosphere and on the smaller scales of electron/ion holes or chains of electron holes, as noted above. Moreover plasma inhomogeneities  and couplings to other effects like impurities should be included, higher order wave and wave-particle interactions and propagation effects. It seems improbable that on the larger scales  quasilinear saturation will play any primary role. We noted the arguments against which are based on the lack of dense enough plasma for radiation trapping. Trapping may, however, become important possibly in holes and result in downward transport of the trapped radiation to low altitudes or even the ground. This trapped radiation will not be subject anymore to saturation because it left its excitation source. All these problems are still badly understood and by no means solved even just in application to AKR. Progress in solving them is highly desirable but requires sophisticated plasma instrumentation in order to resolve the subtleties/fine structure in the energetic electron distribution. It also requires the launch of new spacecraft tangentially traversing the auroral zone at low altitudes of a few 1000 km at most. The ECMI has, according to our judgement, so far proved to probably offer the best model and explanation of then generation of AKR in application to the auroral region, the magnetospheric substorm during magnetospherically disturbed periods as well as during quiet times. During all those times it is most interesting and important to also investigate the relation of AKR to its external source: in the tail this source is with high probability to be found in tail reconnection events. These become violent during substorns/storms in the tail plasma sheet. During times of magnetospheric calm a similar source is located in the distant reconnection region but should be much weaker and close to earth should shift more poleward. On the dayside reconnection at the magnetopause is presumably responsible for AKR generated below the cusp.  

{After all these remarks, it should have become clear that AKR is by far not a resolved problem, even when considering just its role it plays in the magnetosphere as an energetically intense radiation which in the meso- and macrophysics of the magnetosphere is not of primary but high secondary  importance. This importance lies in its diagnostic power in several respects: as a sign of strong non-thermal effects, its relation to non-thermal plasma distributions, its originally unexpected relativistic nature (even though electromagnetic radiation is always tied to relativity) without that it would be of very low intensity only, its challenge to observation and instrumentation, and finally all its listed theoretical challenges. We have focussed mainly on the latter and there just on the most obvious ones, the excitation of the X mode. Observations have indicated that the radiation is not purely in the X polarization; there are contributions of O and also locally Z modes. Even though these are weak, they imply that obliqueness, inhomogeneities in the plasma and magnetic field, reflection, absorption, and possibly even weak nonlinear wave-interactions may contribute to generation of deviations in the polarization properties of AKR. So far they have barely been treated properly and, from the point of view of this note, are lesser dominant in understanding the generation of AKR but just its propagation. This on the other hand is important in a wider view onto space and astrophysics.} Though nowadays there may be more tantalizing questions in magnetospheric physics than solving the large number of open problems of generation of AKR, we repeat that radiation is in a wider astrophysical context the only way of receiving information about distant objects in space. Designing spacecraft and instrumentation as well as improving the theory of generation of radio radiation in a dilute non-thermal plasma serves those needs in an undeniable way simply because there is no other comparably easily accessible laboratory in space with properties like the magnetosphere, in this case the auroral magnetosphere, to investigate and model the processes of intense radio-radiation generation. 

We have listed a number of unsolved physical questions concerning AKR. Fortunately, in the magnetosphere AKR does not provide a substantial energy loss. It is just some moderately intense (actually weak) natural radiation generated in and emitted from the magnetosphere. iI also causes apparently little dissipation globally though locally this point is not clear yet. The reaction of the distribution to its presence has not yet been pinned down even though the mentioned quasilinear depletion is unimportant. Claims that it play an important role (as those we noted have been made in application to the generation of solar radio emission) in limiting its intensity and modifying the electron distribution like whistlers do in the radiation belts, are unrealistic and irrelevant. The plasma is too dilute for such an effect even when the radiation is reflected from the bounds of the auroral cavity on its way out into space. Deformation of the electron distribution is primarily caused by VLF in the magnetosphere, not by AKR \citep{labelle2002}. However, the radiation may be important in its interaction with electron holes if generated by them. This is a most interesting question of relevance both, theoretically and observationally. One of the most interesting recent discoveries is the probable downward transport of trapped AKR, from the source region down into the ionosphere. No really working mechanism has been proposed yet though possibly plasma holes forming caverns when  going down in the downward current region might be capable of trapping some of the radiation as has been speculated \citep{treumann2012a} somewhere else. {Since the ECMI generates radiation below $\omega_{ce}$ which is anyway trapped and unable by itself to escape, this is in principle not unreasonable just requiring that the downward moving electron hole sources may survive down to the ground. Downward moving also implies that they are generated in the downward current region. Altogether this is an effect worth to be highlighted in this context.} Another one is the possible role of upgoing lightnings on the auroral or more generally the upper ionosphere, a problem not yet seriously considered anywhere. These transport comparably large amounts of energy upward from the upper atmosphere into the upper ionosphere. The role they play there has barely investigated yet. 

{In the context of the universe, radiation plays the absolutely decisive role \citep{rybicki1979}. We noted several times already that radiation is the sole source of information we can obtain about distant objects and processes in the distant universe. Unravelling the real mechanism of the generation of AKR assumes enormous importance in this context. This cannot be mentioned often enough. Its non-thermal mechanism indicates the nonthermal and thermodynamically open state of the source region from remote, which otherwise remain inaccessible. The main general importance of AKR  is thus found in its application to the universe. The magnetosphere is the rare if not single case where observations in situ and comparison with theory can be performed at the same time, tested against hypotheses and understood to some quite large extent. Insight into the physical processes related to AKR obtained in situ here has far reaching importance in astrophysical application.} Except in very few cases, radiation from Jupiter, Saturn and possibly from some extra-solar planets, it has been applied to remote objects. However its applicational reservoir in astrophysics is by far not exhausted. The realization that it is based on the electron-cyclotron maser instability gives a mechanism at hand that should be explored in detail in how far it depends on the presence of all those small scale structures in the auroral region in this more general astrophysical view. 

{It is pretty clear, and we close with this remark that, \emph{could we ever perform even one single measurement in situ the remote universe, we would probably have to rewrite much of so far well-accepted, believed and doubtlessly well-established astrophysics}. This was the lesson scientists had always to learn, in particular in near-earth space physics. Any such measurement without any exception provided a wealth of unexpected surprises and enforced re-thinking of  known physics and re-designing well-established models, mostly not completely abandoning but changing and refining them.}

\begin{acknowledgement}
The important contributions over the past half a century by Don Gurnett, Roger Anderson,  Jim LaBelle, Bob Ergun, Bill Kurth, Lou Lee, Philippe Louarn, Don Melrose, Phil Pritchett, Raymond Pottelette, Chung Wu, Peter Yoon, Philippe Zarka and many others of the radio community, whom to list here is impossible, to the research on AKR are gratefully acknowledged collectively. Without all their efforts little progress could have been achieved on the understanding of the deep physical content of AKR which is fundamental to non-thermal radio-radiation, while in magnetospheric physics energetically a marginally important phenomenon only. Some left-out names can be taken from the (as well rather incomplete) reference list. 
\end{acknowledgement}





\end{document}